# Addressing the Dark State Problem in Strongly Coupled Organic Exciton-Polariton Systems

*Evripidis Michail* [1,2], *Kamyar Rashidi* [1,2], *Bin Liu* [2,3,†], *Guiying He* [1,2], *Vinod M. Menon* [1,3], *Matthew Y. Sfeir* [1,2,*]

[1] Department of Physics, Graduate Center, City University of New York, New York, NY 10016, USA

[2] Photonics Initiative, Advanced Science Research Center, City University of New York, New York, NY 10031, USA

[3] Department of Physics, Center for Discovery and Innovation, City University of New York, New York, NY 10031, USA

The manipulation of molecular excited state processes through strong coupling has attracted significant interest for its potential to provide precise control of photochemical phenomena. However, the key limiting factor for achieving this control has been the "dark-state problem", in which photoexcitation populates long-lived reservoir states with similar energies and dynamics to bare excitons. Here, we use a sensitive ultrafast transient reflection method with momentum and spectral resolution to achieve the selective excitation of organic exciton-polaritons in open photonic cavities. We show that the energy dispersions of these systems allow us to avoid the parasitic effect of reservoir states. Under phase-matching conditions, we observe the direct population and decay of polaritons on time scales of less than 100 fs and find that momentum scattering processes occur on even faster timescales. We establish that it is possible to overcome the "dark state problem" through careful design of strongly coupled systems.

* Corresponding Author: msfeir@gc.cuny.edu

† Present Addresses: Department of Electrical Engineering and Computer Science, University of Michigan, Ann Arbor, MI, 48109, USA

There is a broad interest in using strong coupling to control and manipulation of the electronic, optical, and quantum properties of organic materials to enable a wide range of emergent phenomena in optoelectronics, biology, and chemistry, among others.[1, 2] The resulting exciton-polaritons exhibit large oscillator strengths, tunable energies, large delocalization and transport lengths, and enable novel quantum phenomena at room temperature, due to the large exciton binding energies of molecular chromophores.[3, 4, 5] It has been proposed that these engineered excited states will permit access to novel photophysical and photochemical phenomenon by modifying potential energy surfaces in photochemical and isomerization reactions, driving forbidden spin conversion processes, enhancing photoconductivity and light harvesting, as well as enabling the emergence of more exotic phenomena such as polariton condensates.[6,7, 8, 9, 10] However, in practice, much of this potential has been difficult to realize in experimental studies. For example, there have been relatively few successful examples of modified excited state processes in which the primary role of the polariton is to tune an excited state chemical potential that drives the conversion of photons to chemical or electrical energy.[11, 12]

The primary bottleneck for designing novel photochemical systems based on exciton-polaritons has been a lack of a comprehensive understanding of their excited state dynamics which includes the large contributions from dark reservoir states. In strongly coupled systems, ground state optical transitions are dominated by the upper polariton (UP) and lower polariton (LP) states, which are separated by the Rabi splitting energy, $\hbar\Omega_R$, whose magnitude depends on the number of coupled molecules (N) and the dissipation rates.[13] However, the excited state dynamics are strongly affected by the presence of a large density of formally dark "reservoir states" that are uncoupled to the electromagnetic field if the transition energies of all N molecules are assumed to be degenerate. However, in typical organic strongly coupled systems, which feature broad ensemble exciton linewidths and modest Rabi splitting, these states are weakly absorbing and overlap with LP transitions.[15] Similar to bare excitons, the reservoir states are dispersionless and higher in energy than the corresponding LP state.[14] Furthermore, as a result of the direct absorption, large density of states, and long excited state lifetimes of reservoir states, charge or energy transfer processes from strongly coupled molecular systems have been found to largely resemble the uncoupled material.[16, 17] An additional complication is that it is difficult to distinguish between reservoir versus polariton excited-state populations using transient spectroscopy. Numerous experiments have highlighted that photoexcitation of

"reservoir" states modifies strong coupling conditions and obscures the intrinsic polariton dynamics and corresponding transient signals.[18, 19, 20, 21, 22] As such, it is an open question as how to optimize the design of organic photonic systems to enable novel photochemical processes that directly harvest polariton states. This includes persistent uncertainties concerning the rate of coupling between exciton-polaritons and "reservoir" states, the intrinsic lifetime of the polariton and its scattering dynamics, as well as the proper way to measure such interactions in the time domain.[23, 24]

To develop a comprehensive picture of charge carrier dynamics in organic exciton-polaritons systems, concepts related to the characterization of molecular excited states have to be reinterpreted in the context of strongly coupled systems.[16] In molecular systems, excited state dynamics are typically only weakly sensitive to excitation wavelength since the timescale for hot exciton cooling, usually on the order of 1 ps or less, is much smaller than its overall excited state lifetime (~ 1 ns). In contrast, cooling in strongly coupled systems occurs from relaxation of reservoir states to polaritons, which occurs on timescales (~ 100 ns) that are much slower than both the natural lifetime of the exciton (~ 1 ns) and the intrinsic polariton lifetime (< 1 ps). As such, drastically different excited state dynamics should result from different excitation conditions, e.g., reservoir pumping versus direct polariton pumping (**Figure 1a**). Importantly, nearly all previous studies on polariton dynamics have used experimental conditions for reservoir pumping, which precludes an accurate experimental determination of the intrinsic polariton dynamics, including its energy and momentum relaxation processes.[26, 27, 22, 28] As a result, selective excitation of polariton states is essential for studying and optimizing energy and charge transfer phenomena in strongly coupled systems.[29]

Here we show that the dynamics of exciton-polaritons can indeed be directly measured in the time-domain using selective excitation coupled with momentum-resolved detection. Selective excitation of LP states is achieved using photonic modes associated with "open cavity" systems, such as surface plasmon polaritons (SPP) and narrow waveguide modes (WG) (**Figure 1b, c**). The key aspect of these systems is a polarization-dependent, continuous dispersion that extends from low frequency to the exciton resonance, with hybrid exciton-polaritons emerging near the avoided crossing point (**Figure 1c**). Compared to microcavity modes, their smaller effective mode volume, $V_m$, allows us to achieve a large Rabi splitting ($\hbar\Omega \propto (N/V_m)^{1/2}$) with fewer coupled molecules.[30] These characteristics have the dual benefit of increasing the energy spacing between reservoir states and the lower polariton (LP) and decreasing the number of reservoir states.[31] This allows us to achieve selective excitation of LP states by simultaneously selecting for the phase matching condition (PMC) for LP-specific excitation, encompassing the polarization, photon energy, and wavevector that matches the polariton dispersion curve. Using this approach, we find that the intrinsic lifetime of LP states in our structures is < 100 fs and can be measured directly in the time-domain. Finally, we observe a momentum scattering process for selective LP excitation that is much faster than its inherent excited state lifetime.

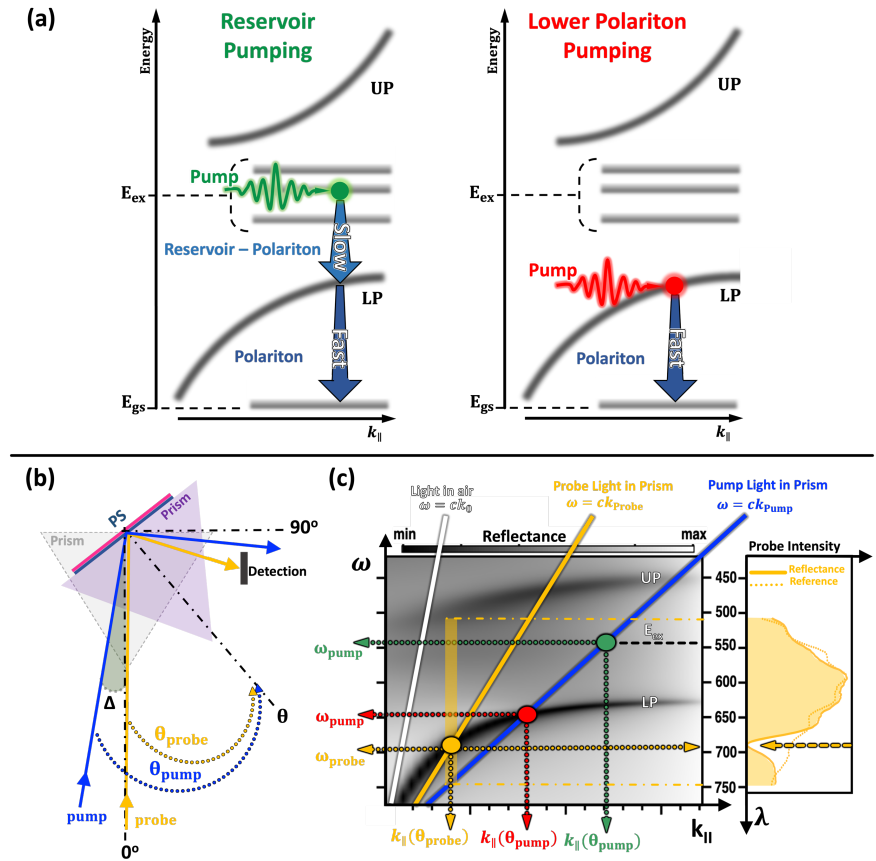

**Figure 1** (a) Schematic energy-momentum diagram, showing the excited state dynamics resulting from reservoir pumping (left, green color) versus direct LP pumping (right, red color). Coupling of reservoir states to LP is a slow process relative to the intrinsic LP decay. (b) Schematic representation of the pump-probe experimental technique in the Kretschmann–Raether configuration, with $\Delta = \theta_{pump} - \theta_{probe}$, that permits selective excitation of LP states. (c) A representative angle-resolved reflectivity map of a strongly coupled "open cavity" system, with key dispersion relations and intersections highlighted. The black dotted line indicates the exciton energy ($E_{ex}$). In pump-probe measurements, the position of the LP resonance at $\theta_{probe}$ is determined from the identification of sharp dips in reflected experimental probe spectrum (right, solid yellow line with arrow) compared to the reference probe spectrum (right, dotted yellow line).

Together, the PMC and large energetic separation between reservoir states and polaritons provided by "open cavity" geometries enables a highly sensitive and selective pump–probe transient spectroscopic method. We have fabricated our open cavity photonic structures (PSs) using a thin film metal-organic stack that exhibit strong coupling between organic molecular exciton and an electromagnetic mode (Methods). The character of these PSs depends on the thickness of the organic layer. For thin layers of e.g., 35 nm we observe electromagnetic modes at the metal–dielectric interface with SPP character (**Figure 2**, upper panel) that exhibits strong coupling to our dye molecule.[30] As the thickness of the dielectric layer is increased to 250 nm ($PS_{WG}$) the field profile is altered. Optical WG modes emerge that exhibit strong confinement within the organic layer, and which can also strongly couple to the dye molecules ($UP_{WG}$ and $LP_{WG}$). This mode exhibits a steep dispersion that is excited by TE polarized light.[32] In addition, we observe a longitudinal SPP-WG mode confined mainly to the metal-organic interface ($UP_{SPP-WG}$ and $LP_{SPP-WG}$) with shallow dispersion that is launched by TM polarized light. Strong coupling is observed for both modes, with a Rabi splitting of $\Omega_{WG} = \Omega_{SPP-WG} = 690$ meV. As the two photonic modes in this system are uncoupled to each other, we observe two distinct curves in measurements of the angle-resolve photoluminescence (PL) (**Figure 2c**), which both show the dispersive behavior associated with exciton-polaritons. As in reflection, the PL dispersion is steeper for the waveguiding mode with emission observed beyond 700 nm, reflecting its significant excitonic character (lower panel, **Figure 2c**).

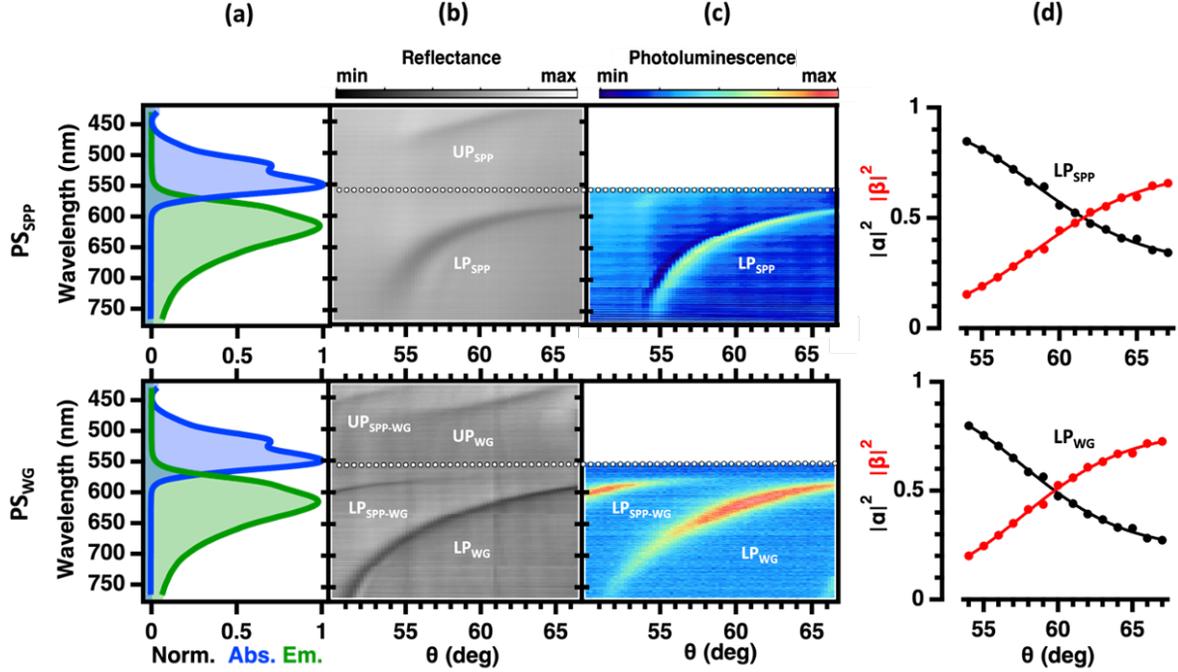

**Figure 2.** (a) The normalized absorption (blue color) and emission (green color) spectrum of the bare organic dielectric film on a glass substrate. Strong exciton-photon coupling in the metal - dielectric photonic structures ($PS_{SPP}$ and $PS_{WG}$) are confirmed through the characteristic anticrossing behavior on the angle-resolved (b) reflectivity and (c) photoluminescence maps. The white dotted lines indicate the exciton energy. The upper panel depicts the metal dielectric photonic structure with surface plasmon polariton character ($PS_{SPP}$), which exhibits strong coupling with two distinct branches ($LP_{SPP}$ and $UP_{SPP}$). The lower panel depicts thicker films support strong coupling to additional modes, including longitudinal SPP-WG modes ($UP_{SPP-WG}$ and $LP_{SPP-WG}$) and transverse waveguide modes ($UP_{WG}$ and $LP_{WG}$). (d) The Hopfield coefficients represent the fraction contribution of photonic $|\alpha|^2$ (black color) and excitonic $|\beta|^2$ (red color) component of the $LP_{SPP}$ (upper panel) and the $LP_{WG}$ (lower panel) for the $PS_{SPP}$ and $PS_{WG}$, respectively.

To disentangle the dynamical evolution of exciton-polaritons from reservoir states, we examined the differential reflectivity ($\Delta R/R$) of our PSs under two regimes of photoexcitation: (i) reservoir pumping where the incident photons are approximately resonant with the bare exciton state and (ii) direct LP pumping with photon energies that are too low to excite the reservoir states directly. To achieve PMC in both linear and transient measurements, we employ the Kretschmann-Raether attenuated total reflection method (**Figure 1b**), which permits excitation of surface waves from the far-field.[33] As depicted in **Figure 1b**, spatially overlapped pump and probe beams propagate through a prism to a thin film stack with an angle relative to its apex of

$\theta_{pump}$ and $\theta_{probe}$, respectively for an angular offset of $\Delta = \theta_{pump} - \theta_{probe}$. We study our PSs in the low excitation regime (< 0.5 µJ/cm$^2$) to minimize photobleaching and other extrinsic effects, e.g., exciton-exciton annihilation, using a high sensitivity transient optical spectrometer that we have recently developed.[34] For direct LP pumping, we restrict our energy ($\hbar\Omega_{pump}$) and momentum ($k_{\parallel,pump}$) to match regions of the polariton dispersion curve for which strong light emission is observed in PL measurements, i.e., states with a significant exciton fraction (**Figure 1c**).

For a pump wavelength resonant with the bare exciton transition at 500 nm, we observe dynamics consistent with the population of long-lived reservoir states. The strongest transient response ($\Delta R/R$) are prominent long-lived features in the spectral region between 620 – 680 nm, coincident with the minimum of the LP$_{WG}$ for the $\theta_{probe}$ ($\lambda_{LP}$ = 650 nm) in reflectivity measurements. (**Figure 3a**). The strong transient response of the LP resonances for reservoir pumping reflects transient changes in strong coupling rather than providing a direct probe of the polariton population. The appearance of the two antisymmetric dispersive features (with maximum $\Delta R/R$ near 640 nm and minimum near 660 nm) at LP resonance energy can be interpreted as a transient blue shift resulting from the photoexcitation of long-lived exciton population which bleaches a fraction of the ground-state absorption (**Figure 3b**).[35] As the Rabbi splitting depends on the square root of the number of coupled molecules, the magnitude of the shift will be a maximum at early times and be reduced as the exciton population decays.[36] This effect can be readily observed in **Figure 3b**, where the zero-crossing point shifts to the red for long delay times as the ground state is repopulated. This characteristic dispersive signal has been reported in previously in cavity-coupled organic systems.[37] As expected, the lifetime of the transient signal for reservoir pumping of the PS$_{WG}$ nearly exactly matches the excited state lifetime of excitons in the bare organic film measured under the same conditions (**Figure 3c** with additional details in **Figure SI–9** of the SI).[38] In both cases, the excited state lifetime can be approximated by a biexponential decay with time constants of 4 ps and 200 ps.

In contrast, direct LP pumping at 630 nm under PMC results in a dramatic shortening of the excited state lifetime compared to the intrinsic exciton lifetime. Data measured at the same $\theta_{pump}$ and $\theta_{probe}$ as reservoir pumping exhibits a similar derivative-shaped transient feature that is centered at $\lambda_{LP}$ = 650 nm (**Figure 3d**). However, features resulting from direct LP pumping exhibit an extremely faster decay and approximately an order of magnitude higher maximum differential reflection signal amplitude in comparison with the signal obtained from reservoir excitation under the same pump fluence (**Figure 3e**). A determination of the excited state is limited by our experimental time resolution limitations, which ranges from 240 – 270 fs going from small to large angles of incidence through the prism (**Figure 3f**).[39, 40] We interpret this fast response as reflecting the intrinsic lifetime of the polariton state based on its ultrafast decay. The expected LP lifetime can be calculated (see section II of the SI) considering the lifetime of the bare cavity, the exciton lifetime, and the Hopfield coefficients.[41] This analysis suggests that the time constant for decay of the LP in $PS_{WG}$ is approximately 40 fs. Furthermore, the order of magnitude increase in signal intensity is consistent with the large absorption cross-section of the polariton compared to reservoir states. Notably, the excited lifetime doesn't change gradually as the pump photon energy is tuned away from the reservoir states, but rather reflects two distinct components whose relative amplitude depends on the excitation wavelength and momentum (**Figure SI–12** at SI).

To further support the assignment that our differential transient signal originates from a real population of LP states, we measure the photoluminescence under the same experimental configuration, including the excitation wavelength, excitation angle, and probe angle (see sec. III of the SI). Indeed, the photoluminescence spectrum features an intense LP emission band at the same wavelength as our strong transient feature in differential reflection measurements (green shaded area in **Figure 3e**). This indicates that our transient data corresponds to a polariton state with significant excitonic character. We further rule out contributions from spurious signals by measuring the differential reflection for different polariton momenta. For example, we have decreased the pump wavelength ($\lambda_{pump}$ = 650 nm) and increased the energy separation between the pump using an angular offset of $\Delta = 6°$ (**Figure 3g**). At this momentum, the LP is more photon-like and the expected lifetime will be decreased to approximately 27 fs. In our measurement, in which our instrument response function is relatively large compared to this value (**Figure 3i**), we observe this change in lifetime as a reduction in the maximum amplitude of the signal compared to a more exciton-like state. Here, our maximum $\Delta R/R$ signal is reduced by a factor of ~ 0.7 for the data in **Figure 3g** compared to **Figure 3d**, fully consistent with the shortening of the excited state lifetime by the same amount.

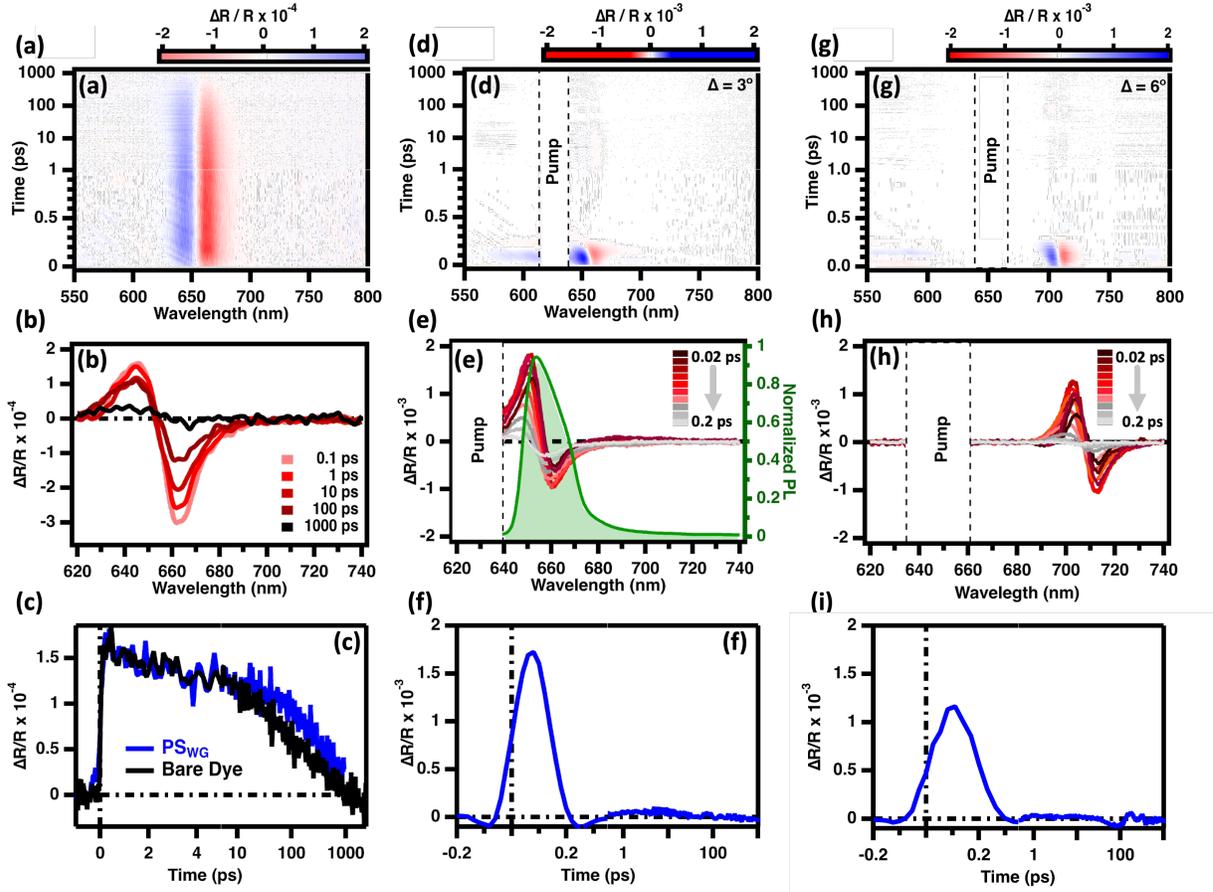

**Figure 3.** Transient reflection measurements on the $PS_{WG}$ demonstrate two different dynamical regimes which can be accessed under different excitation conditions. (a) Raw transient absorption data shows that reservoir pumping at 500 nm (0.5μJ/cm²) with Δ = 3° results in long-lived transient signal that reflects the weak coupling between the reservoir and the polariton states. (b) Transient spectra at key delay times show a dispersive signal associated with the LP state that persists for hundreds of picoseconds. (c) The decay kinetics of the dispersive LP feature (blue line) can be approximated by a biexponential decay with time constants of 4 ps and 200 ps and is nearly identical to the bare dye lifetime measured at the exciton energy (black line). (d - f) The same data for direct LP excitation at $\lambda_{pump}$ = 630 nm (0.5μJ/cm²) with Δ = 3° shows that we can directly excite the $LP^{WG}$ state ($\lambda_{LP}$ = 650 nm) under phase matched conditions. Here we observe only the intrinsic polariton population and decay with a lifetime of less than 100 fs. A strong PL spectrum (green shade) is obtained under the same excitation conditions, indicating that the state has significant excitonic character. In the transient reflection data, regions with strong pump scatter have been excluded. (g - i) The same kinetic response is obtained for a different phase matching condition. Here, $\lambda_{pump}$ = 650 nm, Δ = 6°, and $\lambda_{LP}$ = 710 nm. The greater photonic character of this state results in a shorter lifetime, which causes a reduction in the peak amplitude of our signal.

We map the strong resonant behavior associated with the PMC by comparing the transient differential reflection signal obtained keeping the pump wavelength fixed ($\lambda_{pump}$ = 630 nm) while varying the in-plane momentum of the pump pulse ($k_{\parallel\text{-pump}}$) (**Figure 4**). To ensure consistency between different measurements, we used a constant angular offset ($\Delta \approx 3°$) between the pump and probe beams and a constant pump fluence (0.5 µJ/cm$^2$). Indeed, the magnitude of our transient signal exhibits a sharp decrease as $k_{\parallel\text{-pump}}$ is increased (**Figure 4a**) or reduced (**Figure 4c, d**) compared to optimal PMC at $k_{\parallel\text{-LP res}}$, i.e., $k_{\parallel\text{-pump}} = k_{\parallel\text{-LP res}} + \Delta k_{\parallel}$ (**Figure 4b**), which we attribute a decrease in the maximum LP population resulting from weaker absorption of pump photons. For instance, as $\theta_{pump}$ is increased by 0.7 ° ($\Delta k_{\parallel}$ = 6 x 10$^{-3}$ cm$^{-2}$), the maximum transient reflection signal drops to half ($\Delta R/R$ = 8.4 x 10$^{-4}$) of the maximum value at PMC ($\Delta R/R$ = 1.65 x 10$^{-3}$ at $\Delta k_{\parallel}$ = 0). In this configuration, our pump pulse is red-detuned (solid blue dot in **Figure 4a**) from the polariton state at $k_{\parallel\text{-pump}}$. Similarly, the in-plane momentum of the pump pulse is positively detuned (higher than resonance) compared to the polariton state at $k_{\parallel\text{LP res.}}$ corresponding to the pump energy (horizontal dotted line in **Figure 4a**). A similar effect is observed for blue-detuning in energy (i.e., negative detuning in momentum space), where a momentum change of $\Delta k_{\parallel}$ = − 1.1 x 10$^{-2}$ cm$^{-2}$ reduces the maximum transient reflection signal by one-quarter ($\Delta R/R$ = 4.1 x 10$^{-4}$). Furthermore, at larger values of $\Delta k_{\parallel}$, e.g., $\Delta k_{\parallel}$ = − 1.9 x 10$^{-2}$ cm$^{-2}$, the transient signal is unresolvable. The $\Delta k_{\parallel}$ values for each LP pumping condition were calculated and summarized with appropriate experimental quantities in **Table SI–1** in SI. Significantly, changes in the magnitude of the transient signals, which are proportional to the excited state carrier density, follow the width of the polariton resonances determined using steady-state reflectivity measurements (**Figure SI–14** in SI). A similar dependence of the PL emission intensity on PMC for direct LP pumping is observed (in **Figure SI–15** in SI). This further confirms our interpretation that the observed ultrafast relaxation process corresponds to the intrinsic lifetime of the LP state.

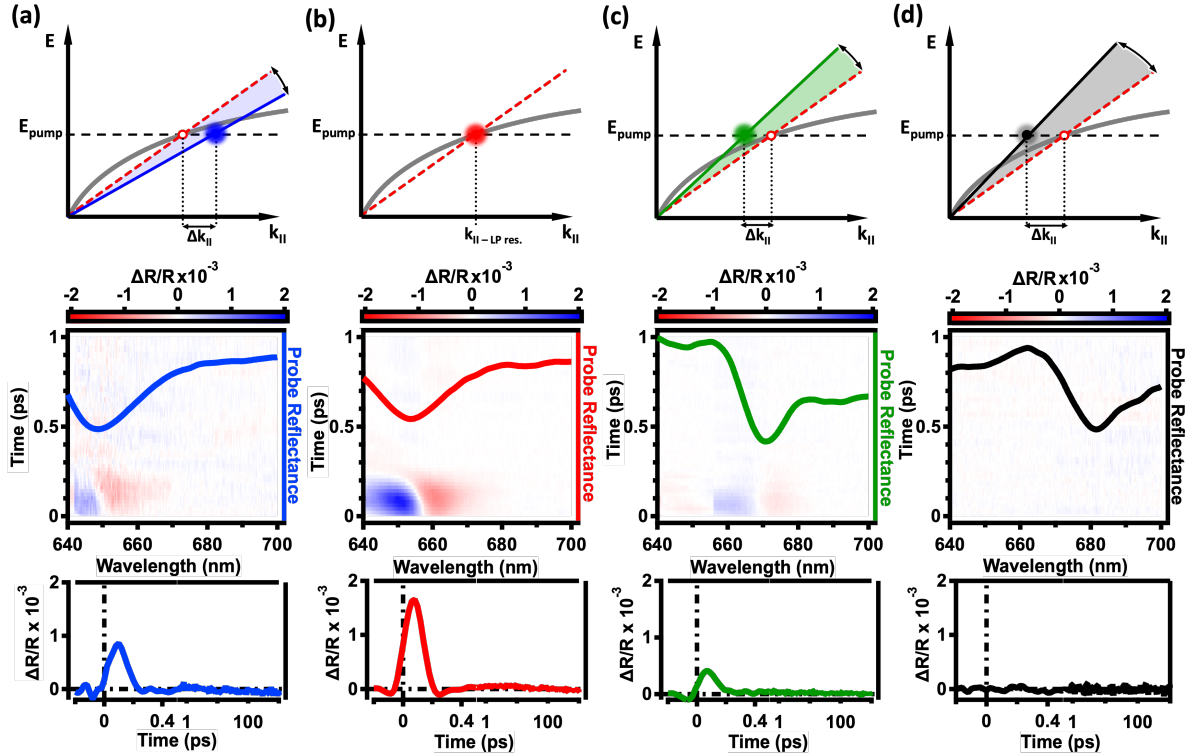

**Figure 4**. For a fixed pump energy, the magnitude of the transient reflection signal from the polariton state is strongly reduced as (a) the in-plane momentum of the pump pulse at $k_{\parallel\text{-pump}}$ is positively detuned ($\Delta k_{\parallel} = 6 \times 10^{-3}$ cm$^{-2}$) from (b) its optimal value for PMC $\Delta k_{\parallel} = 0$). (c) A similar reduction is observed for negative detuning ($\Delta k_{\parallel} = -1.1 \times 10^{-2}$ cm$^{-2}$). (d) If the momentum mismatch is too large ($\Delta k_{\parallel} = -1.9 \times 10^{-2}$ cm$^{-2}$) no transient signal is observed as polaritons are not efficiently pumped. Here the top panel shows a schematic for the experimental configurations in (a-d). The middle panel shows the corresponding raw transient data along with a plot of the steady-state reflectivity with the strong dip indicating the LP resonance. The lower panel shows the kinetic and peak magnitude associated with the LP transient response.

By combining angle-resolved transient reflectivity measurements with angle-resolved PL measurements under PMC, we find that polariton scattering is extremely rapid such that thermalization occurs time scales that are much faster than radiative decay (< 100 fs). This approach allows us to distinguish momentum scattering processes, which change the energy and momentum of the polariton as a function of time, from transient changes in the strong coupling conditions, which are highly sensitive to the population of excited states.[42] For PL measurements, we excited the system at a fixed wavelength and in-plane momentum that is phase matched to the LP state $k_{\parallel\text{-pump}}$ and detect the total integrated emission intensity at distinct momentum vectors

$k_{\|det}$. By systematically increasing the offset between the excitation and detection angle, we measure the efficiency of polariton cooling relative to its intrinsic lifetime, which has an upper bound of < 100 fs. We note that while the experimental momentum resolution for PL (less than 3°) is slightly broader than for pump-probe measurements (less than 1°) due to a larger solid angle for collection, it is considerably narrower than the range of angles probed in the experiment (up to Δ = 9°). Under reservoir excitation, where no momentum conservation is expected, we indeed see that the total integrated intensity is independent of collection angle over the measured range (**Figure 5b**). This is consistent with Fourier space PL measurements (**Figure 2**) and results from a combination of a dimensionless exciton manifold and small scattering rate to LP states that results in a broad distribution of emitting polariton states with distinct momenta. Surprisingly, the same phenomenon is observed for direct LP pumping of a narrow set of polariton momenta. The integrated intensity under pumping conditions of $\lambda_{pump}$ = 630 nm and $k_{\|-pump}$ = 0.86 cm$^{-1}$ ($\theta_{pump}$= 59° ) exhibits a similar insensitivity to collection angle, implying a broad distribution of emitting states (**Figure 5c**). The transient optical and angle-resolved PL measurements together imply that polariton momentum scattering occurs on ultrafast timescales that are much faster the polariton decay (<< 100 fs). The results are consistent with the observation of polariton condensation in organic systems under high pump fluences.[43, 44, 45]

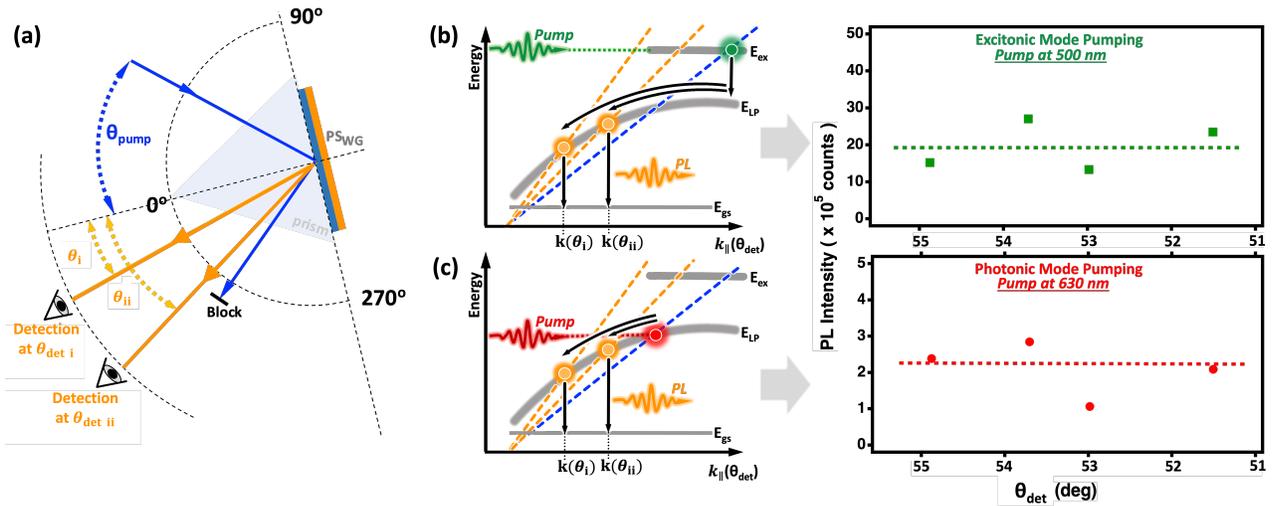

**Figure 5.** (a) Schematic representation of the angle-resolved PL measurements where the PS$_{WG}$ is photoexcited at a fixed wavelength ($\lambda_{pump}$) and in-plane momentum ($k_{\|-pump}$) and the total integrated emission intensity is detected at a series of distinct momentum vectors $k_{\|-det}$. (b) Measurements for reservoir excitation at $\lambda_{pump}$ = 500 nm, $\theta_{pump}$= 59° showing that the total integrated intensity is independent of collection angle $\theta_{det}$. The dimensionless exciton manifold

and small scattering rate to LP states results in a broad distribution of emitting polariton states with distinct momenta. (c) For direct LP excitation under PMC at $\lambda_{pump}$ = 630 nm, $\theta_{pump}$ = 59° the same effect is observed, with the PL emission intensity being insensitive $\theta_{det.}$. The broad distribution of emitting states indicates that polariton momentum scattering occurs on ultrafast timescales that are much faster the polariton decay (<< 100 fs).

The "dark state problem" is persistent and needs to be overcome or avoided through careful design of strongly coupled systems. Dispersion engineering, as is demonstrated here using open cavity systems, is one possible approach to addressing this challenge. Alternatively, systems in the ultrastrong coupling regime may permit large enough energy separations to enable selective excitation, though the polariton eigenstates are further modified. Finally, the dynamics of reservoir states in the condensation regime are not well established, but the emergence of a coherent state may further modify the excited state dynamics. However, even if the effect of reservoir states can be fully mitigated, the explicit charge and energy dynamics of polaritons need to be explicitly measured and optimized. Still, our results suggest that efficient polariton assisted photochemistry is indeed possible in a properly designed system.

**Supporting Information**: The data that support the findings of this study are available from the corresponding author upon reasonable request.

**Notes:** The authors declare no competing financial interest.

**Acknowledgment:** This work was supported by the U.S. Department of Energy, Office of Science, Office of Basic Energy Science under Award Number DE-SC0022036. The authors acknowledge the use of the nanofabrication facility at ASRC. V.M.M. acknowledges funding from the US Air Force Office of Scientific Research MURI (FA9550-22-1-0317). This research used the Advanced Optical Spectroscopy and Microscopy facility of the Center for Functional Nanomaterials (CFN), which is a U.S. Department of Energy Office of Science User Facility, at Brookhaven National Laboratory under Contract No. DE-SC0012704.